# Large Enhancement of Ferro-Magnetism under Collective Strong Coupling of YBCO Nanoparticles

*Anoop Thomas,[†,#] Eloise Devaux,[†] Kalaivanan Nagarajan,[†] Guillaume Rogez,[‡] Marcus Seidel[†,§], Fanny Richard,[†] Cyriaque Genet,[†] Marc Drillon[‡,]\* and Thomas W. Ebbesen[†,]\**

[†]University of Strasbourg, CNRS, ISIS, 8 allée G. Monge, 67000 Strasbourg, France

[‡]University of Strasbourg, CNRS, IPCMS, 23 rue du Loess, 67034 Strasbourg, France

ABSTRACT

Light-matter strong coupling in the vacuum limit has been shown to enhance material properties over the past decade. Oxide nanoparticles are known to exhibit weak ferromagnetism due to vacancies in the lattice. Here we report the 700-fold enhancement of the ferromagnetism of $YBa_2Cu_3O_{7-x}$ nanoparticles under cooperative strong coupling at room temperature. The magnetic moment reaches 0.90 $\mu_B$/mol, and with such a high value, it competes with $YBa_2Cu_3O_{7-x}$ superconductivity at low temperature. This strong ferromagnetism at room temperature suggest that strong coupling is a new tool for the development of next generations of magnetic and spintronic nanodevices.

**Introduction**

Over the past decade, strong coupling has proven to be a promising way to modify and control properties of materials such as charge and energy transport,[1-13] superconductivity,[14-16] work function,[17] non-linear optics[18-21] and chemical reactivity,[22] In strong coupling, the material is typically placed in the confined electromagnetic field of a cavity or a surface plasmon tuned to be in resonance with a material transition. Under the right conditions, new hybrid light-matter states appear, called polaritonic states such as P+ and P- shown in Figure 1, modifying the ladder of energy levels of the material. Furthermore, this occurs even in the dark due to the interaction between the zero-point energy fluctuations of both the optical mode and the material.

Strong coupling is facilitated by collective coupling whereby a large number N of oscillators such as molecules couple to a single optical mode.[23] This enhances the Rabi splitting $\hbar\Omega_R$, the energy that separates the polaritonic states P+ and P-, as √N. In addition, N-1 so-called dark states (DS) are formed. P+, P- and DS are all collective states, delocalized over the optical mode volume (Figure 1A).

In the course of studying the effects of strong coupling of phonons on superconducting microcrystalline powders, we found that while the critical temperature Tc of $Rb_3C_{60}$ is increased by 30 to 45K, that of the $YBa_2Cu_3O_{7-x}$ oxide (YBCO) decreased from 92 to 86K.[14] While the $Rb_3C_{60}$ is a phonon-based superconductor and therefore the change in Tc could be rationalized, the change in YBCO was not expected since the underlying pairing mechanism is not described by BCS theory,[24] and is believed to involve both strong electron correlations and phonon mediated interactions between electrons.[25,26] In trying to understand the change in YBCO using a SQUID magnetometer, we found that in parallel to the superconductivity another property was being modified by strong coupling, namely the magnetic response.

YBCO bulk material is a well-known spin singlet superconductor which exhibits in the normal state antiferromagnetic short-range correlations. However, for nanoparticle (NP) samples, unexpected room temperature ferromagnetism, with a well-defined hysteresis cycle, has been reported which is believed to be related to the presence of oxygen vacancies created at the surface of NPs[27-33]. Such a behavior has given rise to a huge number of studies of oxides, nitrides or sulfides, expected to be non-magnetic but which exhibit room temperature ferromagnetism when they are dispersed at the nanoscale. The net moment is usually very low, and accordingly can be attributed to magnetic impurities, but several reliable experimental and DFT-based theoretical studies conclude that it is likely intrinsic and due to oxygen vacancies.[28,34] For instance, a direct magnetic imaging of the non-magnetic $SrTiO_3$ fully supports this assumption.[31] By using ultrahigh resolution photoemission electron microscopy (PEEM), it was shown that ferromagnetic nanodomains develop at the oxygen-deficient surface of the sample, even for temperatures well above 300K, confirming the existence of intrinsic ferromagnetism.

Here we report, that under collective strong coupling this ferromagnetism is enhanced by two to three orders of magnitude to the point that it even can perturb the superconductivity and lower the Tc of YBCO. There are several parameters that are critical to observe such an enhancement in the ferromagnetism. They can be rationalized by the collective states that favor spin alignment and increase the magnetic domain size.

**Results**

Since YBCO has phonons modes that are weakly absorbing, the best way to strongly couple this material to the surface plasmon mode of Au films is to use the cooperative coupling technique illustrated in Figure 1B. In this approach, the material to be coupled is placed in a polymer matrix which has a strong vibrational band that overlaps one of the YBCO phonon bands. The polymer vibrational band is strongly coupled and transfers the coupling property to the material thru a cooperative process described in detail elsewhere.[35,36] This technique works well for molecular solutions and fine powders as in the present study where the average particle size is around 200 nm (Figure S1A,B). This cooperative regime generates extended states that favors long-distance interaction between particles.

The samples were prepared using commercial YBCO powder purchased from Can Superconductors. X-ray diffraction pattern (see Figure S2) shows that the YBCO has, as expected, the orthorhombic structure (i.e. the 123 phase) and that there is a very small amount of the green phase $Y_2BaCuO_5$. Note that this extra phase does not affect the magnetic properties nor the superconducting transition of $YBa_2Cu_3O_{7-x}$ at 92 K. The YBCO was further ground in a mortar before mixing it with a polymer solution (either polystyrene, PS, or polymethylmethacrylate, PMMA). The mixed solution was then spin coated on sputtered Au film deposited on a high purity Si substrate (the preparation conditions are important, see Supporting Information for more details). The Au film supports the surface plasmons that couple to the material vibrational band (Figure 1B). The FTIR data for the powder used in this study is shown in Figure 1C together with dashed curve. (E) Temperature-dependent magnetization of the YBCO powder (black circles), film of YBCO+PS on Si (black squares), strongly coupled YBCO+PS on Au/Si (red squares) and cooperatively off-resonant YBCO+PMMA on Au/Si (blue squares) in the zero-field-cooled (ZFC; filled squares and circles) and 100 Oe field-cooled (FC; empty squares and circles) modes in the temperature range of 4K to 200K. The onset of superconducting transition (Tc) was determined from the intersection point of the polynomial fits on the ZFC and FC curves. The bare film features a Tc of 92 K, while it is shifted to 87 K for the strongly coupled YBCO as shown by the dashed lines in the respective panels.

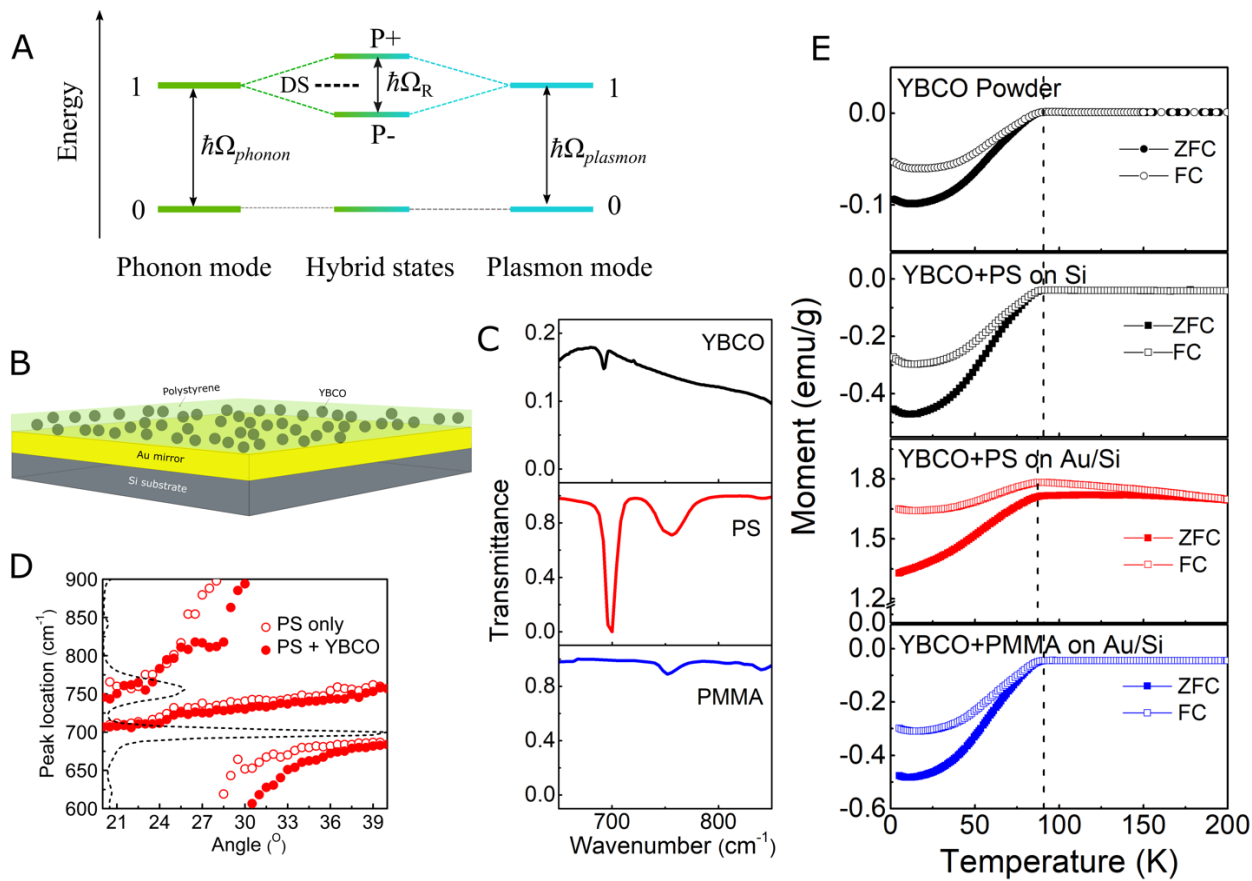

**Figure 1.** (A) schematic illustration of the strong coupling (cooperative) between the phonon mode of the YBCO and the surface plasmon mode of Au film. The hybrid states P+ and P- are separated by the Rabi splitting energy, and the dark states (DS) are represented by the black dashed lines. (B) Cartoonic illustration of the cooperatively strongly coupled YBCO nanoparticles embedded in the polymer matrix. The YBCO containing polystyrene is spin coated on to the Au (10 nm) film sputtered on the Si window. The polystyrene (PS) matrix is shown in green and the embedded YBCO nanoparticles are shown as black circles. (C) FT-IR transmission spectra of the YBCO particles (black curve), PS (red curve) and PMMA (blue curve). (D) Dispersion curves, measured by ATR-FTIR spectroscopy, showing the cooperative strong coupling of the surface plasmon mode with the PS vibration at 697 cm$^{-1}$ which overlaps with the YBCO phonon mode. The empty red circles represent the strong coupling for PS alone, and the solid circles shows the PS+YBCO mixture. The PS transmission spectrum is shown by the black

those of the PS and PMMA. We couple cooperatively the 697 $cm^{-1}$ phonon band of YBCO, using the overlap with the vibrational mode of PS (Figure 1C). The corresponding dispersion curve of the strongly coupled sample with surface plasmon is shown in Figure 1D. PMMA which has no bands that overlap with YBCO phonon mode was used as a reference. The Tc of YBCO only shifts to lower temperatures when the sample is cooperatively coupled using PS, as shown in Figure 1E. In addition to the Tc shift, notice the positive magnetic moment in this condition and the difference between the FC and ZFC curves above Tc, due to the appearance of the enhanced ferromagnetism under cooperative strong coupling.

It should be noted that the IR spectrum of YBCO is very dependent on the origin of the sample and how it was prepared. For instance, the YBCO peak at 697 $cm^{-1}$ that we couple in these experiments has been attributed to an apical oxygen stretching mode of the crystal structure.[36,37] but it could also arise from the small amount carbon retention during the fabrication of YBCO, which generates a barium carbonate ($BaCO_3$) type environment. The latter results from the incorporation of C in the YBCO structure and is mainly associated with the surface as shown in several studies[39,40] and disappears upon high temperature annealing in an inert atmosphere. X-ray Photoelectron Spectroscopy (XPS) confirms that the YBCO in this study contains a small amount of $BaCO_3$ (see Figure S4). Since the Tc of our bare powder is 92 K, it indicates that nevertheless the carbon content is less than 0.04%, according to the literature.[41]

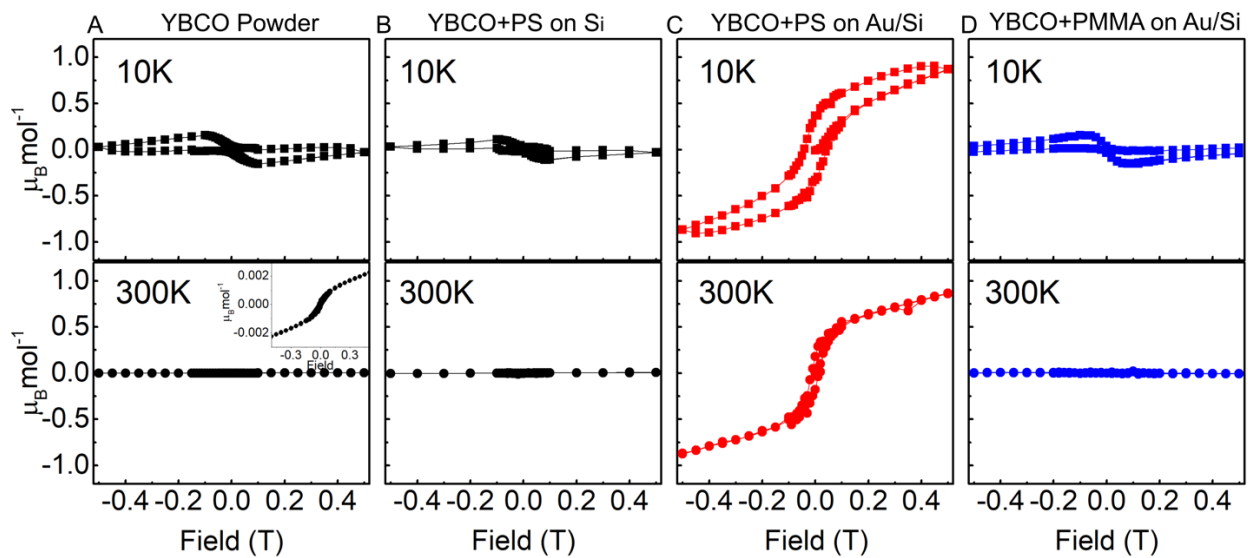

**Figure 2.** M(H) curves of YBCO at 10 K and 300K: (A) YBCO powder; inset in the bottom panel gives the zoom of the M(H) at 300K showing the weakly ferromagnetic character of the powder (B) thin film of YBCO+PS on Si (C) strongly coupled thin film of YBCO+PS on Au/Si and (D) cooperatively off-resonant thin film of YBCO+PMMA on Au/Si. All the curve, except the powder (A), have been corrected for the diamagnetic signal of the Si substrate.

Figure 2 compares the magnetic response of the YBCO at 10K and 300K for various conditions: (A) the original powder, (B) the YBCO dispersed in PS on Si substrate, (C) on Au coated Si substrate, and (D) the same as the latter for YBCO dispersed in PMMA. The M(H) curves at 10K, for samples A, B and D, exhibit the typical butterfly-like behavior of the superconducting YBCO phase. The minimum of magnetization, at 0.1 T, is the critical field beyond which the field penetrates the sample. It agrees well with previous findings which range between 100 and 200mT depending on the material.[41,42] In turn, when using PS as polymer matrix (Figure 2C), a striking change of M(H) occurs, since a ferromagnetic behavior takes place with a saturation value of about 0.90 µB/mol, and a coercive field of 40mT at 10K. At 300K, the pure YBCO powder (sample A) exhibits a paramagnetic behavior with a weak ferromagnetic component detected at low field, but which is much weaker than for the strongly coupled sample (YBCO in PS matrix on Au/Si). Note that the uncoupled samples B and D show at 300K very similar M(H) variations to that of powder sample A.

What is most striking in these experiments is that the magnetic moment is enhanced by a factor of 700 under collective strong coupling at room temperature. The observed magnetic moment of 0.90 $\mu_B$/mol is very large and cannot be explained by assuming that it originates only from anion vacancies at the surface of YBCO NPs since less than 1% of the YBCO unit cells are at the surface. As a result, the contribution of inner NPs oxygen vacancies must be considered. Furthermore, it is not realistic to explain the ferromagnetic contribution at room temperature by the presence of NPs metallic impurities since then their size would have to be several tens of nanometers to remain magnetic at such temperatures. Accordingly, they would be detected by EDAX (Figure S1C), XRD (Figure S2) and XPS (Figure S4, S5), contrary to our findings.

When the same YBCO is dispersed in PMMA on the Au surface, no enhancement of the magnetism is detectable, as expected, since the cooperative coupling is not possible for this polymer confirming the key role of strong coupling in generating the ferromagnetic enhancement.

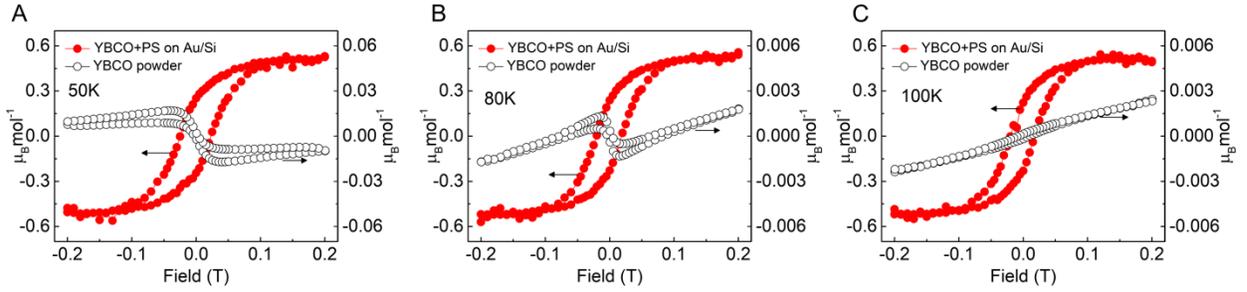

**Figure 3.** (A-C) Comparison of the M(H) curves of the strongly coupled YBCO (YBCO+PS on Au/Si; red filled circles) and YBCO powder (black empty circles) at temperatures near the superconducting transition of YBCO; (A) 50K, (B) 80K and (C) 100K.

The plots of Figure 3 illustrate the striking change of magnetic behavior between uncoupled (powder) and strongly coupled YBCO samples for 3 different temperatures. While the powder sample shows a significant change of M(H) above Tc, the strongly coupled one is dominated by a ferromagnetic-like behavior whatever temperature. At 80K, namely just below Tc, it can be observed that a small ferromagnetic component already exists in the powder sample, which competes with superconducting state, and that this component is multiplied by at least two orders of magnitude in the strongly coupled system.

Since the strong coupling involves the surface plasmons of the Au film, we also studied the effect of the Au film thickness as shown in Figure 4. The hysteretic behavior and, in particular, the net moment of the magnetization increase with the Au thickness and saturates above 100 nm. This is related to the quality of the optical response of the Au film which is known to improve in this thickness range.[44] It confirms the role of the strong coupling induced hybridization between the phonons and the surface plasmons in enhancing the magnetic response.

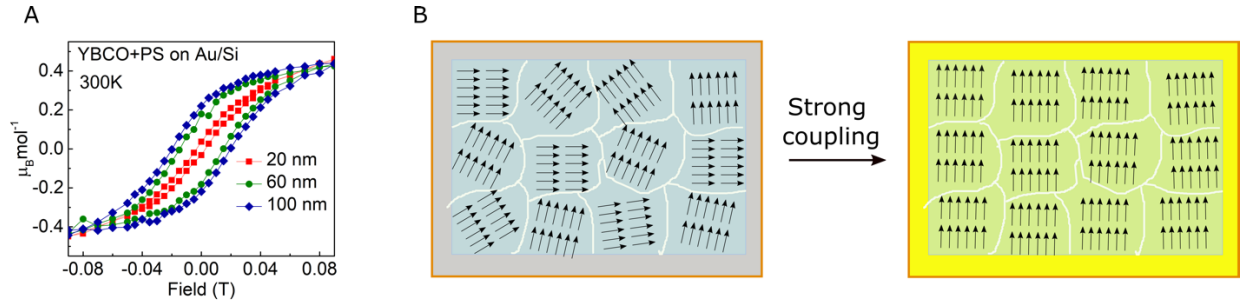

**Figure 4.** (A) M(H) curves of YBCO at 300K for varying Au thickness: 20 nm (red squares), 60 nm (green circles) and 100 nm (blue diamonds). (B) A cartoon of the domain structure of the polymer matrix containing YBCO on Si with the magnetic spins in each domain oriented in differently and the possible alignment of the magnetic spins under strong coupling resulting in the enhanced ferromagnetism.

**Discussion and conclusion**

Similarly to dilute ferromagnetic oxides,[28,45,46] shallow donor electrons, related to oxygen vacancies, form magnetic polarons that mediate magnetic ordering. However, both the very high ordering temperature, above 300K, and the significant magnetic moment (0.90 $\mu_B$/mol) make such a model unrealistic to explain our results. Note that this also rules out the effect of impurities made of transition metal oxide or pure metal nanoparticles. As a result, surface states alone can no longer explain the magnetic behavior and a model based on the effect of a strong coupling between plasmons, generated in the bulk YBCO compound, and spin carriers must be considered.

Ferromagnetic domains already exist in pure nanosized YBCO powder, due to oxygen vacancies, and it has been well-established that they can coexist with superconductivity when the magnetic field is less than a threshold value H* beyond which Cooper pairs are destroyed.[29,30]

These magnetic domains would be related to a hoping mechanism of the unpaired electrons like that reported for dilute magnetic semiconductors. Such a mechanism is well known in mixed valence systems where the transfer of an excess electron between neighboring sites is allowed only for parallel spin configurations of the valence electrons.[47-49] Then, ferromagnetism is promoted. The magnitude of the coupling being given by the transfer integral (of the order of 103K), such that very high ordering temperatures are expected, as observed for YBCO NP powders ($T_{Curie}$> 300K).

Under strong coupling, the behavior of YBCO NPs exhibits striking variation as evidenced by the magnetic moment that is boosted by orders of magnitude and remains so even at room temperature. Then, the competition with the superconducting states below Tc appears crucial because of the antagonism between both properties. A description of the behavior necessitates introducing the coupling between electronic states and phonon modes of the network. At very low temperature, the growth of magnetic domains is most likely collective and related to the breaking of Cooper pairs for an increasing external field. The delocalization of the unpaired electrons then promotes nanodomains to line up in the field direction which in turn reinforces existing ferromagnetic domains. In this process, coherent phonon modes induced by the strong coupling play a major role. In turn, when decreasing field below H*, the spins condense into Cooper pairs progressively so as to form the superconducting nanodomains but this recombination is limited by the internal field of the ferromagnetic domains.

At high temperatures, only ferromagnetic domains remain. The variation of the net moment at 300K, from 0.037 to 0.22 µB/mol for Au thickness ranging from 20 to 100 nm, confirms the key role of plasmon resonance in the growth of magnetic domains. Furthermore, the value of the magnetic moment points out that the inner spins participate in the magnetic ordering and is not limited to the NPs surface.

Finally, the existence of ferromagnetism at room temperature makes such systems suitable for the development of next generations of spintronic nanodevices. Usually, the limitation in NPs magnetic materials stems from the fact that they are characterized by low temperature blocking temperature. Strong coupling of NPs materials such as YBCO may be promising solution to get around this limitation. Our results, together with the recent theoretical prediction of enhanced ferro-electric phase transition,[50] further broadens the appeal of strong coupling with the vacuum field to engineer material properties.

**Supporting Information**.

Experimental methods, FT-IR, SEM, EDAX, XRD and XPS characterization of YBCO powder

AUTHOR INFORMATION

**Corresponding Authors**

* Marc Drillon, University of Strasbourg, CNRS, IPCMS, 23 rue du Loess, 67034 Strasbourg, France. marc.drillon@ipcms.unistra.fr

* Thomas W. Ebbesen, University of Strasbourg, CNRS, ISIS, 8 allée G. Monge, 67000 Strasbourg, France. ebbesen@unistra.fr

**Present Address**

[#]Department of Inorganic and Physical Chemistry, Indian Institute of Science, Bengaluru, 560012 - Bengaluru, India.

[§]Deutsches Elektronen-Synchrotron DESY, Notkestraße 85, 22607 Hamburg, Germany.

**Author Contributions**

The manuscript was written through contributions of all authors. All authors have given approval to the final version of the manuscript.

**Notes**

The authors declare no competing financial interest.

ACKNOWLEDGMENT

We thank Stefan Reisner and Tobias Adler (Quantum Design Europe, Darmstadt) for their help in SQUID magnetometer measurements. We acknowledge support of the International Center for Frontier Research in Chemistry (icFRC, Strasbourg), the ANR Equipex Union (ANR-10-EQPX-52-01), the Labex NIE Projects (ANR-11-LABX-0058 NIE), CSC (ANR-10- LABX-0026 CSC), and USIAS within the Investissement d'Avenir program ANR-10-IDEX-0002-02, the ERC (project no 788482 MOLUSC) and QuantERA project RouTe. M.S. acknowledges support from the Marie Skłodowska-Curie actions of the European Commission (project 753228, PlaN).

*Supporting Information for*

# Large Enhancement of Ferro-Magnetism under Collective Strong Coupling of YBCO Nanoparticles

*Anoop Thomas,[†,#] Eloise Devaux,[†] Kalaivanan Nagarajan,[†] Guillaume Rogez,[‡] Marcus Seidel,[†,§] Cyriaque Genet,[†] Fanny Richard,[†] Marc Drillon[‡,]\* and Thomas W. Ebbesen[†,]\**

[†]University of Strasbourg, CNRS, ISIS, 8 allée G. Monge, 67000 Strasbourg, France

[‡]University of Strasbourg, CNRS, IPCMS, 23 rue du Loess, 67034 Strasbourg, France

[#]Present address: Department of Inorganic and Physical Chemistry, Indian Institute of Science, Bengaluru, 560012 - Bengaluru, India.
[§]Present address: Deutsches Elektronen-Synchrotron DESY, Notkestraße 85, 22607 Hamburg, Germany.

1- **Methods**
2- **SEM characterization of YBCO powder**
3- **XRD characterization of YBCO powder**
4- **FTIR spectrum of YBCO powder**
5- **XPS analysis of YBCO powder**

## 1 - Methods

The superconducting powder YBa$_2$Cu$_3$O$_{7-\delta}$(YBCO) was purchased from Can Superconductors and used without further purification. The powder was ground well in a mortar before use. The polymers, polystyrene (PS, analytical standard for GPC; MW 200,000) and polymethyl methacrylate (PMMA, MW 120,000), and their solvent, toluene were purchased from Sigma Aldrich/Merck. Toluene was distilled and dried before use. Silicon (Si) wafers (4-inch diameter, 500 μm thick) purchased from TEDPELLA were cut into 2x2 cm pieces and cleaned by sonication in water and subsequently in isopropanol. The cleaned Si wafers were dried in a hot air oven and used as substrates for thin film making. The thin films of YBCO nanoparticles embedded in different polymers were prepared as follows. The YBCO powder (50 % by wt. compared to the polymer) was added to freshly prepared homogeneous solutions of polymer (PS or PMMA, 20 wt%) and sonicated for 4 minutes (20 kHz, 10 second ON - 5 second OFF sequence) and then stirred for at least six hours at room temperature. Metal (Au) deposition onto the cleaned Si substrates was done by sputtering technique using Emitech K575X sputterer (Parameters: Current 60 mA; the Au thickness was varied from 20 nm to 100 nm by changing the sputtering time). The films (4 μm thick) were prepared by spin casting (RPM: 1000, time: 2 minutes) the YBCO-polymer mixture onto Si substrates (for reference samples) and Au coated Si substrates (for strongly coupled samples). The samples were stored in a vacuum box before measuring the magnetic properties. To measure the magnetization properties of the samples, the spin casted films were cut into small pieces (ca. 4x4 mm) : few of them were then inserted into a gelatine capsule, sealed using Kapton tape.

The temperature-dependent magnetization (ZFC and FC) at 100 Oe magnetic field and the field-dependent magnetization measurements were done in ISIS, Strasbourg (using MPMS SQUID magnetometer), in IPCMS, Strasbourg and Quantum Design Europe, Darmstadt (using MPMS3 SQUID magnetometer). The infrared spectra were recorded using Bruker Vertex70 FTIR spectrometer, in transmission and ATR mode. The dispersion curves of the strongly coupled samples were measured in ATR mode with a Variable Angle Reflection Accessory ("Bruker" A513/Q). For the dispersion measurements, a right-angle ZnSe prism, sputter-coated with 10 nm of Au and spin-coated with a thick YBCO embedded polymer layer (5 μm) at the bottom, was placed at the sample position. The angle of incidence on the prism was varied from 20º to 40º with steps of 0.5º for both TM and TE polarizations. By recording the reflection of an Au film sputter-coated at the bottom of the ZnSe prism, the instrument response of the reflection accessory was measured and subtracted from the sample data. Scanning Electron Microscopy (SEM) characterization was done using SEM model Quanta FEG 250, FEI company), equipped with an Energy-Dispersive X-ray detector (EDX, EDAX company). X-ray Photoelectron Spectroscopy was achieved using a Thermo Scientific K-Alpha XPS system. The AlKα source produces X-rays (hv = 1486.7 eV). The level of vacuum was $10^{-8}/10^{-9}$ mbar in the main chamber. The spot size of the X-ray beam was fixed at 400 µm. The YBCO powder was also characterized by X ray diffraction to analyze carefully its composition. The PXRD measurements were carried out on a D8 Discover_Bruker powder diffractometer in the Bragg-Brentano geometry. The diffractometer was equipped with a front monochromator (Cu Kα1 wavelength λ = 0.154 056 nm) and a LynxEye XE-T linear detector.

## 2 – SEM characterization of YBCO powder

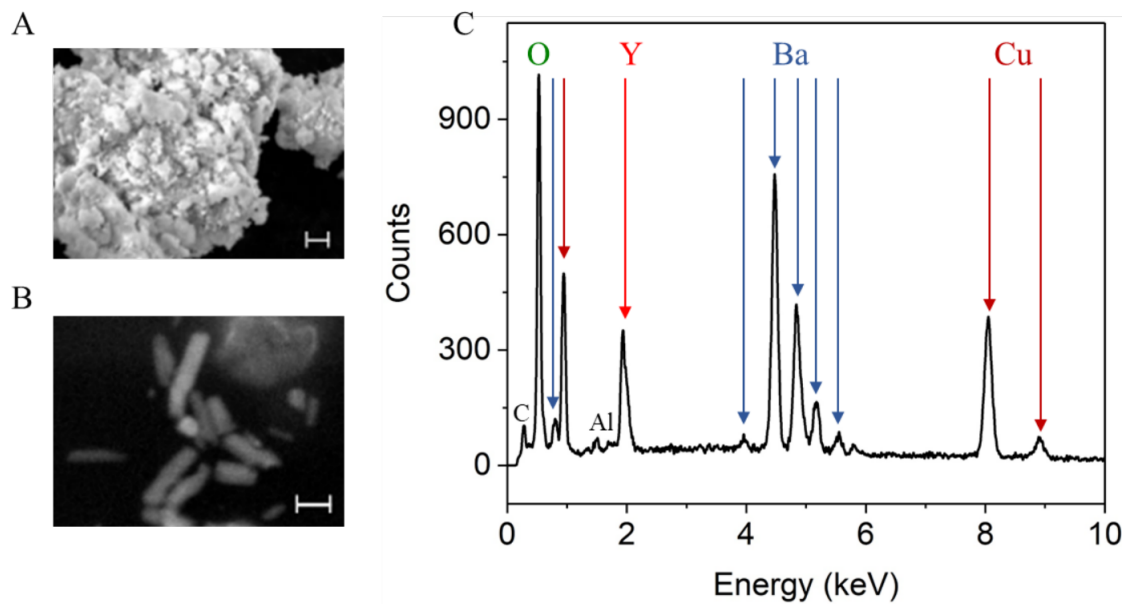

**Figure S1.** A and B are SEM images of the YBCO powder. C is an EDX spectrum of the same powder. Clusters of powder can have a size as large as 10µm (image A, scale bar is 1µm) but are composed of smaller nanoparticles having an average size around 200nm (image B, scale bar is 100nm). EDX spectra were also recorded (image C) and did not show any impurity into the powder (Al and C elements are coming from the sample holder and adhesive tape used to fix and hold the powder into the vacuum chamber of the SEM).

## 3 – XRD characterization of YBCO powder

The YBCO powder was also characterized by X ray diffractometry to analyze carefully its composition. Initial powder showed two main phases: Y123 and Y211 as well as a few others phases (Copper Oxide, Barium Copper Oxide, etc.). Note however that the critical temperature of this powder (Tc=92K) is not affected by those different phases and that they are just due to the preparation method of the powder.

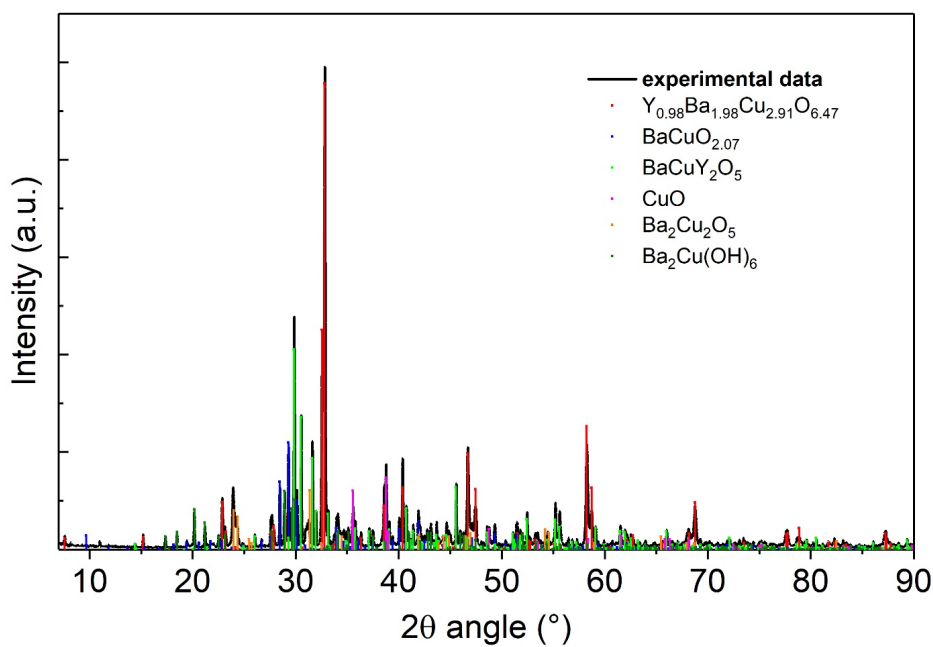

**Figure S2.** XRD diagram of the YBCO powder. Full black line: experimental data. Color vertical lines correspond to phase identification (in red and bright green are the Y123 and Y211 phase respectively, blue, pink, orange and olive green correspond to copper oxide, barium copper oxides and barium copper hydroxide). Numbers in the figure caption refer to the JCPDS number for each phase.

**4– FTIR spectrum of YBCO powder**

Pellets of KBr ground with YBCO powder were prepared (15 mg of YBCO in 100 mg of KBr) and measured. The peaks at 697 cm$^{-1}$, 846 cm$^{-1}$ and 1400 cm$^{-1}$ are typical of $CO_3$ (see also Ref. 1)

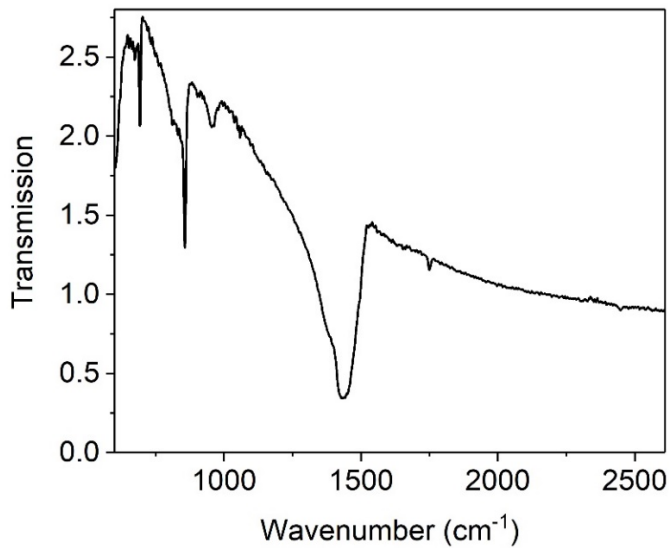

**Figure S3.** FTIR spectrum of a pellet of YBCO/KBr showing peaks typical of carbonate ions.

**5 – XPS analysis of YBCO powder**

XPS was used to analyze the YBCO powder andto estimate its $CO_3$ content. First the survey only showed the presence of expected elements plus carbon, that is always found in samples, proving one more time the purity of the powder in terms of composition (Table S1).

**Table S1**: Survey of the elements detected by XPS, with the peak position (in eV) and atomic content (% of the total area probed).

| Name | Peak Binding Energy (eV) | Atomic % |
|---|---|---|
| Ba $3d_5$ | 780.09 | 10.83 |
| O $1s$ | 531.21 | 53.1 |
| Cu $2p_3$ | 934.04 | 9.5 |
| Y $3d$ | 158.4 | 3.14 |
| C $1s$ | 285.37 | 23.43 |

Then following the orbitals C $1s$ and Ba $3d$, we confirmed the presence of carbonate into the powder. There is a clear signature of $CO_3$ in the C $1s$ fitting (peak at 289eV; Ref. 2). In addition, the fitting of the Ba $3d$ spectrum reveals that most of the $Ba^{2+}$ ions contained in the powder are bonded to $CO_3$ (peaks at 780eV and 795eV) because we see no trace of the pure $Ba^{2+}$ orbital at 778eV.

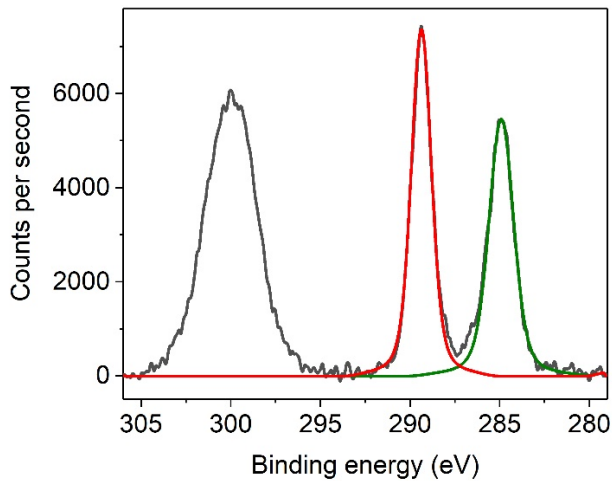

**Figure S4.** XPS spectrum (in black) of the YBCO powder: the peak at 289eV is fitted by $CO_3$ (in red) and the peak at 285eV is fitted by C-C (in green) and is typical of adventitious carbon. The peak at 300 eV corresponds to the orbital Y$3p_{3/2}$.

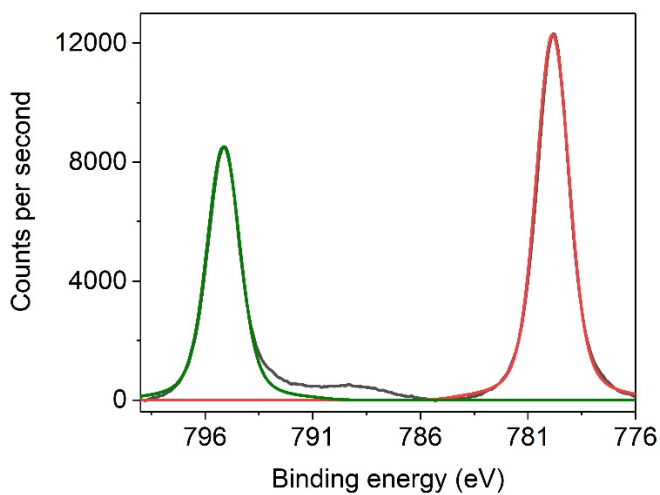

**Figure S5.** XPS spectrum (in black) of the YBCO powder: the peaks at 780eV and 795eV (doublet) are fitted by $Ba^{2+}$ bonded to $CO_3$ (in green and red). The doublet of pure $Ba^{2+}$ ions should stand at slightly lower energies (i.e. 778 and 793eV).